\documentclass[aps,pre,showpacs,amsmath,amsfonts,superscriptaddress,twocolumn]{revtex4}
\usepackage{graphicx}
\usepackage{dcolumn}
\usepackage{bm}
\usepackage{pslatex}
\begin{document}

\title{Quantum Monte Carlo simulation in the canonical ensemble at finite temperature}

\author{K.~Van~Houcke}
\author{S.M.A.~Rombouts}
\author{L.~Pollet}
\affiliation{Universiteit Gent - UGent, Vakgroep Subatomaire en
  Stralingsfysica \\
  Proeftuinstraat 86, B-9000 Gent, Belgium}
\date{\today}

\begin{abstract}
A quantum Monte Carlo method with a non-local update scheme is
presented. The method is based on a path-integral decomposition and a worm operator
which is local in imaginary time. It generates states with a fixed number of
particles and respects other exact symmetries.  Observables like the equal-time Green's
function  
can be evaluated in an efficient way. To demonstrate the
versatility of the method, 
results for the one-dimensional Bose-Hubbard model and a nuclear
pairing model are presented. Within the context of the Bose-Hubbard model the efficiency of the algorithm is discussed. 
\end{abstract}


\maketitle

\section{Introduction}

Quantum Monte Carlo (QMC) simulation is a powerful and versatile
method for the investigation of thermodynamic properties of many-body
systems. When generating a Markov-chain of configurations using
a Metropolis scheme~\cite{Metropolis}, it is known that updates based on local changes
are inefficient, particularly near critical points. At transitional points this
type of algorithm gives very large autocorrelation times, a phenomenon one
refers to as 'critical slowing down'~\cite{vonderlinden}. By developing
non-local update schemes, this problem has been overcome for second order phase transitions.
The Wolff single cluster algorithm~\cite{Wolff} and the
Swendsen-Wang multiple cluster algorithm~\cite{Swendsen}, both used to solve
classical physics problems, were successful applications of this idea.
In the same spirit, loop
algorithms~\cite{Evertz, Evertz03, Kawashima} were developed
for the study of discrete quantum systems. New configurations are obtained by
flipping clusters in the form of loops. 
The systematic error caused by the  
the Suzuki-Trotter approximation~\cite{Suzuki,
  Trotter} was eliminated by formulating the 
world-line algorithms directly in continuous imaginary time ~\cite{Beard, Prok}. 
In the worm algorithm~\cite{Prokofiev}, 
the partition function is embedded in an extended configuration space, allowing a direct and exact evaluation of the one-body Green's
function.  
 The concept of non-local loop updates has also been implemented in  
the Stochastic
Series Expansion (SSE) representation~\cite{Sandvik}, leading to 
'operator loop' update algorithms~\cite{Sandvik99, Dorneich} and
'directed loop' algorithms~\cite{Syl02, Syl03}, 
which are a further optimization of the loop construction. 
The general idea behind this is to construct moves in a
locally optimal way~\cite{Pollet04}. 

All the non-local update world-line algorithms which are mentioned above
sample the
grand-canonical ensemble. In this way one can generate configurations with e.g. varying
magnetization or occupation number. 
Results for the canonical ensemble are obtained by using only the
configurations with the right particle number \cite{Roos99} or by rejecting loop updates
which change this number \cite{Evertz03}. It is clear that this is an inefficient way of
working. 
Sampling the canonical ensemble directly would be advantageous when 
studying systems 
where particle-conservation is important.
One example is
the transition between the superfluid and the Mott phase in the Bose-Hubbard model
at constant filling. This transition belongs to the 
(d+1)-dimensional XY 
universality class \cite{Sachdev}. 
When entering the superfluid phase, it becomes
difficult to keep the number of bosons constant and tuning the
chemical potential can become a very time-consuming task. 
When simulating mesoscopic systems
like superconducting grains \cite{Dukelsky99, Delft01} or atomic nuclei
\cite{Koonin97, Rombouts98}, 
it is primordial  to keep the number of particles constant. 
A world-line algorithm satisfying these conditions is presented \cite{Letter} and discussed
in detail in this paper.
Besides particle-number conservation, the algorithm allows to include other symmetry-projections.
It is constructed in such a way that all moves are accepted, which
makes it efficient to run and easier to code. 
Though working in the canonical ensemble, our
algorithm is still able to generate configurations with different winding numbers, 
in contrast to the local world-line method by Hirsch {\it et al.} \cite{Hirsch81}.
We will demonstrate the versatility of the method by applying it to bosons and to
paired fermions.

\section{The algorithm}

Practically all QMC methods are based on a decomposition of
the evolution operator $e^{-\beta H}$. The trick is to break up this operator
into pieces which can be handled exactly \cite{Rombouts03}.
Generally one can write the Hamiltonian as consisting of an easy part $H_0$
and a residual interaction $V$,
\begin{equation}
  H= H_0 - V.
 \label{eq:hamiltonian}
\end{equation}
The minus sign has been included to ease notations further on.
For such a Hamiltonian, one can make an exact perturbative expansion in $V$ of the evolution
operator, using
the following integral expression~\cite{Prok}:
\begin{eqnarray}
 e^{-\beta H} 
     & = & \sum_{m=0}^{\infty} 
    \int_{0}^{\beta} d t_m \int_0^{t_m} d t_{m-1} \cdots  \int_0^{t_2} d t_1  
     V(t_1)\nonumber\\
    & &   V(t_2)  
 \cdots   V(t_m)
       e^{-\beta H_0},
       \label{eq:decompos}
\end{eqnarray}
with $V(t) = $exp$(-tH_0) V $exp$(tH_0)$ and $\beta$ the inverse temperature (also called imaginary time).
The basic idea of the continuous-time loop algorithm \cite{Beard, Evertz03}
and the worm algorithm \cite{Prokofiev} is
to insert two
adjoint world-line discontinuities.
By propagating one of these
discontinuities (which are called the mobile 'worm head' and stationary 'worm
tail') through lattice space and imaginary time, 
the configuration changes in such a way that detailed balance is
fulfilled. At that point one is not sampling according to the partition
function Tr$(e^{-\beta H})$,
but according to an extension
hereof,
\begin{equation}
Tr(W^{\dag} e^{-\tau H} W e^{-(\beta - \tau)H}) ,
\label{eq:tradworm}
\end{equation}
with $\tau$ the imaginary time interval between the worm 'head' operator
$W^{\dag}$ and 'tail' operator $W$. The worm head can be creating or
annihilating, depending on the choice of $W$. 
After some Markov steps, the worm head bites its tail and the discontinuities
are removed. 
Only configurations with continuous world-lines can contribute to
the statistics according to Tr$(e^{-\beta H})$.
In contrast to these algorithms, we will work with a
worm which is local in imaginary time.
The evolution operator extended by such a local worm (an imaginary
time-independent operator $A$ to be defined below) reads
\begin{equation}
  U'(\beta,\tau) = e^{- \tau H} A e^{- (\beta -\tau) H},
  \label{eq:evolution}
\end{equation}
where $\tau$ can be regarded as the
imaginary time at which the worm operator is inserted.
We will show that by working with a local worm operator one can construct a
very efficient sampling method,
provided that the worm operator commutes with the residual interaction
($AV=VA$). If $A$ furthermore commutes with the generators of a symmetry of
$H_0$ and $V$, one can restrict the sampling to configurations with those
specific symmetries, leading to symmetry-projected results. In particular one
can sample the canonical ensemble with a worm operator that conserves particle number.
We would like to emphasize at this point that the algorithm stated below
does not depend on the specific structure of $A$.
The operator $A$ has to be chosen in such a way that an ergodic Markov
chain can be
constructed, and therefore it will depend on the specific form of the interaction $V$.

If one decomposes the trace (restricted to the wanted particle number and
symmetry) of
$U'(\beta,\tau)$ using Eq. (\ref{eq:decompos}) and inserts complete
sets of eigenstates of  $H_0$ at all imaginary times, one obtains a set of integrals
which can be evaluated using Monte Carlo sampling. The Markov process will sample
the configurations proportional to the weights
\begin{eqnarray}
 & & W(m,i,t,\tau) =
  \langle i_0 | V | i_1 \rangle e^{- (t_2-t_1) E_{i_1}} 
  \langle i_1 | V | i_2 \rangle e^{- (t_3-t_2) E_{i_2}} 
     \cdots \nonumber \\
  & &    \langle i_{L-1} | V | i_L \rangle e^{-(\tau-t_L) E_{i_L}} 
           \langle i_L | A | i_R \rangle e^{-(t_R-\tau)E_{i_R}}
       \langle i_R | V | i_{R+1} \rangle 
     \cdots \nonumber \\
  & &     \langle i_{m-1} | V | i_m \rangle e^{- (t_m-t_{m-1}) E_{i_{m}}}  
       \langle i_{m}   | V | i_0 \rangle e^{- (\beta + t_1-t_m) E_{i_0}}.
  \label{eq:wormdeco}
\end{eqnarray}
Each configuration is specified by an order $m$ (the number of
interactions), a set $i$ of
eigenstates  of $H_0$ (with $i$ a shorthand notation for all the intermediate states $|i_0 \rangle,
\ldots, |i_L\rangle, |i_R\rangle, \ldots ,|i_m \rangle$), 
interaction times $t_1, \ldots, t_m$,
and the worm insertion time $\tau$. 
We use the notation $E_{i_j} = \langle i_j|H_0|i_j\rangle$.
The configuration $|i_L \rangle$ to the left of the worm is changed by the
worm operator into the configuration $|i_R
\rangle$.
We use the subscript $L$ ($R=L+1$) to indicate the eigenstate $|i_L\rangle$ ( $|i_R\rangle$)
and
interaction time $t_L$ ($t_R$)
just before (after) the worm operator in imaginary time.
We will call the configurations for which $i_L=i_R$ diagonal configurations.
By choosing the worm operator such that its diagonal elements are constant
(i.e. $\langle i|A|i \rangle = c$ for all basis states $|i \rangle$), the sum of the
weights of all diagonal configurations is proportional to 
the particle-number projected trace of the evolution operator $U(\beta)$.
This is nothing else than the 
canonical
partition function Tr$_N(e^{-\beta H})$, with Tr$_N$ the particle-number projected
trace.
Hence, sampling the configurations proportional to their weights $W(m,i,t,\tau)$
leads to a sampling of the canonical ensemble.
The Markov process is set up 
using the Metropolis-Hastings algorithm \cite{Metropolis,Hast70}, hereby sampling
in an extended space according to Tr$_N[U'(\beta,\tau)]$.
At each Markov step only a few of the factors of Eq. (\ref{eq:wormdeco}) are
altered by the worm operator which moves to a new point in imaginary time. These worm
operator moves will be constructed in a such a way that detailed balance is
fulfilled locally at each Markov step. 
Therefore it is also
fulfilled when going from one diagonal configuration to another.
It takes a number of Markov steps before diagonal observables
(i.e. observables which commute with $H_0$) can be
measured again. While each Markov step contains only local changes, the chain of
steps between two diagonal configurations corresponds to a global update.
Non-diagonal operators can be measured using the fact that one
samples in an extended space. By keeping track of the worm moves between two
diagonal configurations, statistics for the expectation values of non-diagonal
operators can be collected, similar to the way one evaluates the one-body
Green's function in the worm algorithm \cite{Prokofiev}. Our method however
will lead to much  better statistics for equal-time non-diagonal
operators, because the worm operator is always local in imaginary time. 

\begin{figure}[h]
\begin{center}
\includegraphics[angle=0, width=7.5cm] {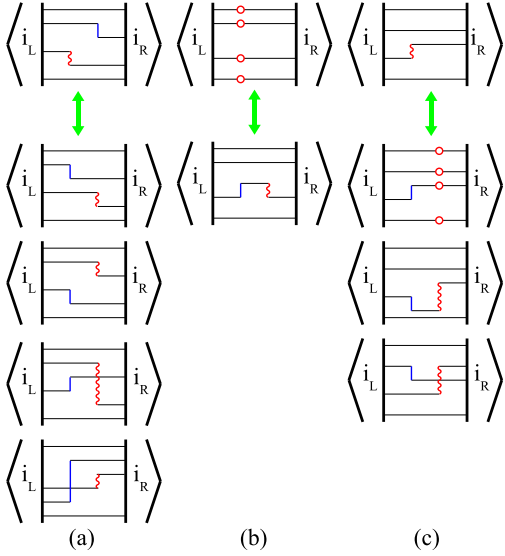}
\caption{ A diagrammatic representation for one-body worm operator moves. The
  worm operator is represented by a curly line and the interaction
  $V$ by a solid vertical line. We distinguish between the following updates. (a) When the worm operator moves in the $D=R$
  direction, it can encounter some interaction. The worm operator can pass the
  interaction, in that way changing the intermediate state. For general
  one-body worm and interaction operators, there are at most four
  possible ways in doing this.  (b)-(c)
At the beginning of the worm move, we introduce the possibility of inserting
  an interaction. When the initial worm is diagonal (represented by circles), a number of interaction
  insertions of the type shown in (b) are possible. In part (c) the initial worm
  operator is not diagonal, and an interaction insertion can make the
  worm diagonal or not. 
   \label{fig:diagram}}
\end{center}
\end{figure}

Before shifting the worm operator over some imaginary time interval $\Delta \tau$, a direction $D$ has to be chosen. 
One can choose between the directions $D=R$ (higher values of $\tau$) and $D=L$
(lower values of $\tau$) with some probability $P(D)$, to be specified later.
The presence of the exponentials in Eq. (\ref{eq:wormdeco}) inspires us to
choose the time shift $\Delta\tau$
proportional to an exponential distribution,
  \begin{equation}
    P(\Delta \tau) d \Delta \tau = \varepsilon_{D} e^{-\varepsilon_{D} \Delta \tau } d \Delta \tau,
    \label{eq:poisson}
  \end{equation}  
with $\varepsilon_D$ an optimization parameter. In shifting the worm operator from
$\tau$ to a new imaginary time $\tau' = \tau + \Delta \tau$, the worm operator can encounter an
interaction operator $V$ at some time in between. Assume the direction $R$ is chosen and the worm
operator meets an
interaction at time $t_R$.
Let us first consider the situation where the worm operator moves through this interaction,
without annihilating it,
\begin{equation}
\langle i_L | A | i_R \rangle \langle i_R | V | i_{R+1} \rangle
\longrightarrow  \langle i_L | V | i'_R \rangle \langle i'_R | A | i_{R+1}
\rangle .
\label{eq:movethrough}
\end{equation}
When passing the interaction, the intermediate state can change.
A
convenient way to pick one of these possible changes is to choose the new configuration proportional to its
weight 
\begin{equation}
  P_{R+1,L}(i'_R) = \frac{\langle i_L | V| i'_R \rangle \langle i'_R | A | i_{R+1} \rangle} 
  {\langle i_L | V A | i_{R+1} \rangle}.
  \label{eq:transv}
\end{equation}
Part (a) of Figure \ref{fig:diagram} shows a diagrammatic representation of the different
ways in which a general one-body worm operator $A=\sum_{i,j} c_{ij} a^{\dag}_i
a_j$ (for some constants $c_{ij}$ ) can
pass a similar interaction $V$. The worm operator is represented by a curly line and the interaction by
a vertical straight line. For this type of worm and interaction operator there are always at most
four ways in which the intermediate state can change.
It should be noted that our choice Eq. (\ref{eq:transv}) is not unique and possibly more optimal
choices can be found \cite{Pollet04}.
Because of this choice however, there appears a factor
\begin{eqnarray}
   n_{R+1,L} & = & \frac{\langle i_L | V | i'_R \rangle \langle i'_R | A |
     i_{R+1}\rangle P_{L,R+1}(i_R) }
{\langle i_L | A | i_R \rangle \langle i_R | V | i_{R+1} \rangle  P_{R+1,L}(i'_R)}   \nonumber \\
& = & \frac{\langle i_L | V A | i_{R+1} \rangle}{\langle i_L | A V | i_{R+1} \rangle}, 
\end{eqnarray}
in the acceptance factor of the Metropolis-Hastings algorithm. Every time the worm operator
passes an interaction an analogous factor appears, depending on the direction
$D$ of propagation.
Therefore it's advantageous to impose on $A$ the
condition,
\begin{equation}
AV-VA = 0, 
\label{eq:commutator}
\end{equation}
because then $n_{DD'} = 1$ in all cases, and we do not have to worry about this
normalization factor anymore. 
Furthermore, Eq. (\ref{eq:commutator}) ensures that the worm operator can
always pass the interaction it encounters. If one would choose a worm operator
$A$ that does not
satisfy this condition, as in grand-canonical algorithms, 
it is possible the worm operator cannot pass the interaction and the direction
of propagation has to change, in that way undoing changes previously made. 
It is intuitively clear that these 'bounces' give rise to a slow decorrelation and
should be avoided \cite{Syl02, Syl03, Pollet04}.
In the directed loop algorithm 
one increases the efficiency of the loop update
by minimizing the appearance of
bounces, but they cannot be eliminated completely because of the reversibility
condition. 
In what follows we will assume the condition
Eq. (\ref{eq:commutator}) is fulfilled, making the algorithm bounce-free. We will drop the factors $n_{DD'}$ to
ease the equations. 
After passing through the interaction at time $t_D$, 
one has to choose a new imaginary-time shift $\Delta \tau$.
However, one can avoid generating a new random number 
by taking the new shift as follows:
\begin{equation}
  \Delta \tau = (\tau'- t_D) \frac{(\varepsilon_{D})_{\rm{old}}}{(\varepsilon_{D})_{\rm{new}}},
\label{eq:deltatau2}
\end{equation}
where the parameter $\varepsilon_D$ has been updated after passing the interaction.

The choice of
the parameters $\varepsilon_D$ follows from detailed
balance.
Because the time shifts $\Delta\tau$ of the worm operator are chosen by Eqs. (\ref{eq:poisson}) and
(\ref{eq:deltatau2}), adding the constraint
\begin{equation}
  E_R-E_L = \varepsilon_L - \varepsilon_R,
\label{eq:condeps}
\end{equation}
ensures
 that all the
 exponentials which appear in the acceptance factor of the Metropolis-Hastings algorithm
 cancel, leading to an efficient sampling method. So in practice one can choose
 any positive value for $\varepsilon_L$ and $\varepsilon_R$, as long as
 Eq. (\ref{eq:condeps}) is fulfilled at each step of
 the worm movement.
To conclude, we write down the acceptance factor for the above procedure when
the worm operator does not change the number of interactions,
\begin{eqnarray}
q & = & \frac{W(m,i',t',\tau') P(i', t', \tau' \rightarrow i,t,\tau)}{W(m,i,t,\tau) P(i,t,\tau
  \rightarrow i', t', \tau')}    \nonumber \\
  & = & \frac{(\varepsilon_{D'})_{\rm{initial}}}{(\varepsilon_D)_{\rm{final}}},
\label{eq:accepvconst}
\end{eqnarray}
where $(\varepsilon_D)_{\rm{final}}$ ($(\varepsilon_{D'})_{\rm{initial}}$) is the value of
$\varepsilon_D$ ($\varepsilon_{D'}$) at the
end (beginning) of the worm operator move into direction $D$, and $D'$ denotes
the opposite direction of $D$.
The actual acceptance
probability is given by $\rm{min}(1,q)$, according to the Metropolis-Hastings algorithm.

Let us now introduce a number of steps, 
which allow 
to change 
the number of interactions in a reversible
way. 
We want the acceptance factor of such updates to be local,
i.e. the probability to pass, create or annihilate an
interaction should only depend on the properties of the state at that point in
imaginary space-time. 
We consider three extensions of the procedure outlined above where
no interactions are created or deleted: 
\begin{itemize}
\item At the beginning of the Markov step, we introduce the possibility that the worm operator creates a new interaction
with probability $c_D$, which depends on the direction $D$ of propagation. This creation will also change the intermediate
state: 
\begin{equation}
\langle i_L | A | i_R \rangle
\longrightarrow  \langle i_L | V | i' \rangle \langle i' | A | i_{R}
\rangle, 
\label{eq:creation}
\end{equation}
assuming again the worm operator is moving in the $D=R$ direction. 
The new intermediate state $|i' \rangle$ will be chosen with probabilities
\begin{equation}
  P_{RL}(i')   =   \frac{\langle i_L | V| i' \rangle \langle i' | A | i_{R}
  \rangle}{\langle i_{L} | V A| i_{R} \rangle}. 
\label{eq:insertprob}
\end{equation}
For a worm operator move in the $D=L$ direction, probabilities $P_{LR}(i')$ can be
defined in an analogous way. 
Figure
\ref{fig:diagram} (parts (b) and (c))  shows a diagrammatic
representation of the insertion of a one-body interaction at the beginning of
the worm move. 
For a diagonal configuration 
only the diagonal part of $A$ contributes to the
matrix element $\langle i_{L} | A | i_{R} \rangle$.
In this situation  the worm operator is represented by little circles and all
world-lines are continuous.  
We will call this
the 'diagonal worm'.
\item When the worm operator arrives at an interaction, one also has to consider the possibility of
annihilating that interaction. Suppose the interaction can be deleted. Let
$a_D$ be the probability to annihilate the interaction while continuing the
worm update, and $s_D$ the probability to annihilate the interaction and stop the
worm update. Then $1-a_D-s_D$ is the probability to pass through that interaction and
continue the worm operator.
\item 
To maintain reversibility, one also has to include the possibility that the
construction of the Markov step does not halt at the moment the worm operator
has finished a shift $\Delta\tau$ without encountering an interaction. At that
point one has to choose between stopping the worm operator, or to continue, with the
possibility of inserting a new interaction at that point. Let $f_D$ be the
probability to continue the worm operator without inserting an interaction, $g_D$ the
probability to insert an interaction and to continue the worm operator, then
$1-f_D-g_D$ will be
the probability to stop the worm operator, without inserting an interaction.
\end{itemize}
After creating, annihilating or passing an interaction, a new time shift
$\Delta \tau$ should again be chosen according to Eqs. (\ref{eq:poisson}) or (\ref{eq:deltatau2}).
Note that the parameter $f_{D}$ is redundant:
jumping with a parameter $\varepsilon_D$ and continuing the worm
operator unaltered with probability $f_{D}$
is statistically equivalent to making a jump with parameter $ \varepsilon_D (1-f_D)$,
and then choosing between either stopping the worm operator or inserting an
interaction and move on.
Therefore, one can set $f_{D}=0$ without loss of generality.
We now determine how the other parameters should be chosen in order to
satisfy detailed balance.
Assume a direction $D$ is chosen.
When no interaction is inserted at the beginning of the worm move, a factor
\begin{equation}
q_{D}^{0} = \frac{\varepsilon_{D'} (1-g_{D'})}{1-c_D},
\label{eq:q^0}
\end{equation}
appears in the acceptance factor. If on the other hand an interaction is
created at the initial time $\tau$ of the worm operator, this will
lead to a factor
\begin{equation}
q_{D}^{c} = \frac{\mathcal{N}_{DD'} s_{D'}}{c_D},
\label{eq:q^c}
\end{equation}
with
\begin{equation}
    \mathcal{N}_{DD'} =  \frac{\langle i_{D} | V A| i_{D'} \rangle}{\langle i_{D} |A| i_{D'} \rangle}.
\label{eq:N}
\end{equation}
A new intermediate state is chosen with probabilities Eq. (\ref{eq:insertprob}).
At the end of the Markov step, the worm operator will annihilate an
interaction or not, leading to extra factors in the global acceptance factor which have the
inverse form of Eqs. (\ref{eq:q^c}) and (\ref{eq:q^0}), because of the inverse symmetry
between beginning and end of the move.
At intermediate points, we can encounter the following situations. 
The worm
operator can
stop after a shift $\Delta\tau$ between two interactions, insert an interaction
and move on. The inverse situation of
this can also occur, when an interaction is annihilated and the worm operator moves
on. Or the worm operator can simply
pass an interaction without annihilating it.   
In order to have a good total acceptance factor, we will require that these
intermediate steps do not contribute to the acceptance factor. This condition leads to
the constraints
\begin{eqnarray}
\mathcal{N}_{DD'} a_{D'} & = & \varepsilon_{D} g_D, \\
a_D + s_D & = & a_{D'} + s_{D'}.
\label{eq:as}
\end{eqnarray}
Apart from that, we want the sampling to be as uniform as possible, which
suggests the condition $q_{D}^{0} = q_{D}^{c}$. 
Putting all this together, the total acceptance factor is given by
\begin{eqnarray}
q & = & \frac{W(m',i',t',\tau') P(m', i', t', \tau' \rightarrow m,i,t,\tau)}{W(m,i,t,\tau) P(m,i,t,\tau
  \rightarrow m',i', t', \tau')}       \nonumber \\
& = &  \frac{(q_D)_{\rm{initial}}}{(q_{D'})_{\rm{final}}},
\label{eq:totalaccep}
\end{eqnarray}
where
\begin{equation}
q_D = \varepsilon_{D'} + \mathcal{N}_{DD'} (s_{D'} - a_D).
\label{eq:q_D}
\end{equation}
The factor $(q_D)_{\rm{initial}/\rm{final}}$ has to be evaluated at the beginning and the end of the Markov step in direction $D$
(with $D'$ the opposite of $D$). The creation probability is given by Eq. (\ref{eq:q^c}), 
\begin{equation}
c_D = \frac{\mathcal{N}_{DD'} s_{D'}}{q_D}.
\label{eq:c_D}
\end{equation}
We still have to determine how to choose the direction $D$.
The acceptance ratio of Eq. (\ref{eq:totalaccep})  inspires us to 
choose the direction of the move with
probabilities
\begin{eqnarray}
P(D=R) & = & \frac{q_R}{R_{LR}}, \\
P(D=L) & = & \frac{q_L}{R_{LR}}.
\label{eq:Pdir}
\end{eqnarray}
with $R_{LR} = q_R + q_L$. 
By accepting all worm operator moves a distribution given by 
\begin{equation}
W'(m,i,t,\tau) = R_{LR} W(m,i,t,\tau), 
\label{eq:qW}
\end{equation}
will be sampled, instead of the distribution $W(m,i,t,\tau)$.
Because the factors $R_{LR}$ fluctuate only mildly in practice, accepting all
moves still leads to a 
a very useful sampling method.
It speeds up the algorithm and reduces the complexity of the code.
Finite
temperature observables can still be evaluated by taking the extra weighting
factor into account: 
\begin{eqnarray}
\langle Q \rangle_{\beta} & = & \frac{Tr[e^{-\beta(H_{0}-V)}
  Q]}{Tr[e^{-\beta(H_0-V)}]} \nonumber \\
& = & \frac{\sum_{s \in \mathcal{S}} (Q_s/(R_{LR})_s)}{\sum_{s \in \mathcal{S}} (1/(R_{LR})_s)}.
\end{eqnarray}

Each time the worm operator creates, annihilates or passes an interaction, the
 parameters $\varepsilon_D$, $c_D$, $a_D$, $s_D$, $g_D$ are determined by
 Eqs. (\ref{eq:condeps}), (\ref{eq:as}),
 (\ref{eq:q_D}) and (\ref{eq:c_D}). This still leaves a lot of freedom. 
We will focus here on two limiting cases. 
First we will consider the case where one of the two parameters $\varepsilon_R$
and $\varepsilon_L$ is always zero. 
In that way it can occur the time shift 
$\Delta \tau$ becomes infinite. This amounts to the worm operator directly jumping to the next
interaction, which speeds up decorrelation in imaginary time direction. 
In order to obtain a worm move
that changes the configurations as much as possible, the parameters
$g_L$, $g_R$, $a_L$ and $a_R$ are maximized. 
The set of parameters obtained in this way is shown in Table \ref{table:solA} for
 the case $E_R > E_L$. 
We will call this solution A. 
The solution for the case $E_L>E_R$ can be found by interchanging the subscripts $L$ and $R$. 
Note that in this solution the worm operator always starts to move into the direction of the
highest diagonal energy.
It is
clear that whenever the worm operator is moving in the direction of the highest
diagonal energy or whenever $E_R=E_L$, the time shift $\Delta\tau$ becomes infinite. 
There are a number of extra conditions one should keep in mind.
Assume the worm operator starts to move in the direction $D=R$ (because $E_R>E_L$). When the worm
operator arrives at an interaction that can be annihilated, one has to
determine $E_L$, $E_R$ and $\mathcal{N}_{LR}$ after the annihilation. If $E_R>E_L$ is still
satisfied then, $s_R$ and $a_R$ from Table \ref{table:solA} are the correct
probabilities to stop or continue the worm operator. If now  $E_L>E_R$ on the other
hand, one has to use the solution $s_R = \rm{min}(1, \frac{E_L-E_R}{\mathcal{N_{LR}}})$ and $a_R
= 0$, but the worm operator keeps moving in the
same direction. The time shift of the worm operator is only
finite when it moves in the direction $D$ and $E_{D} < E_D'$.
Note also that only $g_L$ is mentioned in Table \ref{table:solA}, because
the parameter $g_D$ has only meaning when the time shift is finite. 
In the present solution however, a problem arises whenever $E_L = E_R$. In
this case $\varepsilon_R = \varepsilon_L =0$ and the time shift $\Delta \tau$
is always infinite. Because $s_R = s_L = 0$ in addition, the worm operator never
halts. As a consequence configurations with a diagonal worm will never be sampled.
This can be solved by proposing a small but finite stopping probability. 
This alternative solution for the case $E_L=E_R$ is also given in Table \ref{table:solA}.
The global parameter $\phi$ should be taken small (such that $0< \phi
<\mathcal{N}_{DD'}$ for all diagonal configurations) but not zero, 
and can
be chosen in order to optimize the decorrelation between
successive evaluations of observables. 
Note $R_{LR}$ of Eq. (\ref{eq:qW}) takes the constant value
 $2 \phi$.

\begin{table}
\begin{tabular}{|c||c|c|}
\hline
\hline
parameters & diagonal configurations & diagonal configurations \\
 & $(i_L = i_R)$ & $(i_L \neq i_R)$ \\
 & $(E_L = E_R)$ & $(E_L < E_R)$ \\
\hline
\hline
$\varepsilon_R$ & 0 & 0 \\
$\varepsilon_L$ & 0 & $E_R-E_L$ \\
$q_R$ & $\phi$ & $E_R-E_L$ \\
$q_L$ & $\phi$ & 0  \\
$c_R$ & 1 & min$(1, \frac{\mathcal{N}_{LR}}{E_R-E_L})$ \\
$c_L$ & 1 & 0 \\
$s_R$ & $\frac{\phi}{\mathcal{N}_{LR}}$ & 0 \\
$s_L$ & $\frac{\phi}{\mathcal{N}_{LR}}$ & min$(1, \frac{E_R-E_L}{\mathcal{N}_{LR}})$ \\
$a_R$ & 0 & min$(1, \frac{E_R-E_L}{\mathcal{N}_{LR}})$ \\
$a_L$ & 0 & 0 \\
$g_L$ & 0 & min$(1, \frac{\mathcal{N}_{LR}}{E_R-E_L})$ \\
$R_{LR}$ & $2\phi$ & $E_R-E_L$ \\
\hline
\hline
\end{tabular}
\caption{\label{table:solA} 
A set of algorithm parameters satisfying Eqs. (\ref{eq:condeps}), (\ref{eq:as}),
 (\ref{eq:q_D}) and (\ref{eq:c_D}) for the cases $E_L = E_R$ and $E_L < E_R$
 (otherwise interchange the subscripts $L$ and $R$). We call this solution A, for which one of
 the parameters $\varepsilon_{D}$ is always zero.}
\end{table}

Another possibility to find algorithm parameters follows from the
idea that we want the step size $\Delta \tau$ to be of the order of the mean
imaginary time interval between two interactions.
Therefore we consider the case where 
one of the two parameters $\varepsilon_R$
and $\varepsilon_L$ is $\mathcal{N}_{LR}$. As a consequence the time
shift is always finite. 
For $E_R > E_L$ such a set of parameters is given in Table \ref{table:solB}. 
Again, the case $E_L=E_R$ needs an alternative solution, since otherwise $R_{LR}$
would be zero for diagonal configurations. We will refer to this solution as
solution B.

\begin{table}
\begin{tabular}{|c||c|c|}
\hline
\hline
parameters & diagonal configurations & diagonal configurations \\
 & $(i_L = i_R)$ & $(i_L \neq i_R)$ \\
 & $(E_L = E_R)$ & $(E_L < E_R)$ \\
\hline
\hline
$\varepsilon_R$ & $\mathcal{N}_{LR}$  & $\mathcal{N}_{LR}$ \\
$\varepsilon_L$ & $\mathcal{N}_{LR}$  & $\mathcal{N}_{LR} + E_R-E_L$ \\
$q_R$ & $\mathcal{N}_{LR}$ & $E_R-E_L$ \\
$q_L$ & $\mathcal{N}_{LR}$ & 0  \\
$c_R$ & $\phi$ & 0  \\
$c_L$ & $\phi$ & 0 \\
$s_R$ & $\phi$ & 0 \\
$s_L$ & $\phi$ & 0 \\
$a_R$ & $\phi$ & 1 \\
$a_L$ & $\phi$ & 1 \\
$g_R$ & $\phi$ & 1 \\
$g_L$ & $\phi$ & $\frac{\mathcal{N}_{LR}}{\mathcal{N}_{LR} + E_R - E_L}$ \\
$R_{LR}$ & $2\mathcal{N}_{LR}$ & $E_R-E_L$ \\
\hline
\hline
\end{tabular}
\caption{\label{table:solB} 
An alternative set of parameters for which one of the parameters $\varepsilon_D$
is always $\mathcal{N}_{LR}$. We
call this solution B. The parameter $\phi$ can be chosen to optimize
the algorithm.
}
\end{table}

In short, the algorithm is based on a time-dependent perturbation in the interaction $V$
(see Eq. (\ref{eq:decompos})), so there is no systematic error arising from
time discretization. Because we choose time shifts of a worm operator
according to Eq. (\ref{eq:poisson}) the diagonal part of the Hamiltonian is handled exactly.
There are a number of
algorithm parameters, which have to satisfy Eqs. (\ref{eq:condeps}), (\ref{eq:as}),
 (\ref{eq:q_D}) and (\ref{eq:c_D}).
We have derived two sets of algorithm parameters, satisfying these equations. In the first set (solution A)
one of the parameters $\varepsilon_D$ is always zero and in the second
 (solution B) it is equal to  $\mathcal{N}_{LR}$.  Therefore the main difference
 between these two  solutions will be the size of the imaginary time shift
 $\Delta \tau$.
Other algorithms where
$\varepsilon_R$ or $\varepsilon_L$ take values between $0$ and $\mathcal{N}_{LR}$
can be constructed in a similar way, taking for now only these limiting cases. 
In the next section our QMC algorithm  will be applied to the
Bose-Hubbard model. The efficiency of the two different solutions leading to different algorithms
will be compared in this context.

\section{Application to the Bose-Hubbard model} 

Ultracold bosonic atoms in an optical lattice 
are described by the Bose-Hubbard model \cite{Jaksch, Greiner, Fisher},
\begin{equation}
H = -t \sum_{\langle i,j \rangle}^{N_s} b^{\dagger}_i b_j + \frac{U}{2} \sum_{i}^{N_s} n_i(n_i-1),
\end{equation}
with $b^{\dagger}_i$ ($b_i$) the boson creation (annihilation) operator on
site $i$, $n_i$ the number operator on site $i$ and $\langle i,j \rangle$ denoting nearest neighbors.
The lattice has $N_s$ sites, occupied by $N$ bosons.
The parameter $t$ is the tunneling amplitude and $U$ is the on-site repulsion
strength.
We will restrict the discussion here to the one-dimensional homogeneous case
without trap. 
At low values of $U/t$ the system forms a compressible superfluid. 
This phase is characterized by a gapless excitation spectrum and long-range
phase coherence. 
By
increasing $U/t$, a quantum phase transition
from a superfluid state to a Mott insulating state is achieved at integer
densities. In the pure Mott phase the bosons are localized at individual
lattice sites and all phase coherence is lost due to quantum fluctuations. 
In addition, density fluctuations disappear when entering the Mott phase
and a gap appears in the excitation spectrum.
This phase driven transition 
belongs to the Berezinskii-Kosterlitz-Thouless (BKT)~\cite{Berezinskii70, Kosterlitz73} universality class in one
dimension. 
The Bose-Hubbard Hamiltonian can be rewritten in the form 
Eq. (\ref{eq:hamiltonian}),
\begin{eqnarray}
H_0 & = & \frac{U}{2} \sum_{i} n_i(n_i-1), \nonumber \\
V & = & t\sum_{\langle i,j \rangle} b^{\dagger}_i b_j.
\label{eq:bhham}
\end{eqnarray}
As argued above, it is advantageous to take the worm operator $A$ such that it
commutes with $V$. An operator that satisfies this condition is
given by 
\begin{equation}
A = \frac{1}{\bar{N}} \sum_i n_i + \sum_{i\neq j}  b^{\dagger}_i b_j, 
\label{eq:bhworm}
\end{equation}
with $\bar{N}$ a c-number to be optimized.  In our calculations this
parameter is always set equal to the total number of particles. We have checked our code by
comparing with exact diagonalization results for small lattices.
Ergodicity was tested numerically.
Power law behavior of correlation functions coincides with predictions from
bosonization theory for large lattices.

\begin{figure}[h]
\begin{center}
\includegraphics[angle=270, width=8.5cm]{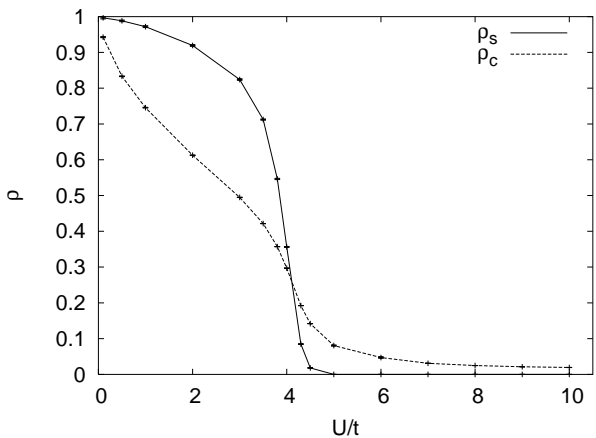}
\caption{Superfluid ($\rho_s$) and condensed fraction ($\rho_c$) for the
  one-dimensional homogeneous
  Bose-Hubbard model as a function of $U/t$. 
The fractions have been calculated for a uniform lattice with $N_s = 128$ 
at an inverse temperature $\beta = 128 t^{-1}$, using Eqs. (\ref{eq:cf})
and (\ref{eq:sf}). The condensed
fraction was calculated from the correlation function $\langle b^{\dagger}_i
  b_{i+r} \rangle$, for which we have high statistics. 
   \label{fig:rho}}
\end{center}
\end{figure}

The one-body Green's function
$G(r) = \langle  b^{\dagger}_i b_{i+r}  \rangle $ is calculated by updating
the entry $r$  in a histogram for $G(r)$ at every Markov step. 
The function $G(r)$ can be normalized directly from the diagonal /
non-diagonal worm fraction.
The non-diagonal worm components $b^{\dagger}_i b_{i+r}$ of
Eq. (\ref{eq:bhworm}) can be given a different weight, leading to a worm matrix
representation of the symmetric Toeplitz type (i.e. a symmetric matrix with
constant values along negative-sloping diagonals). Such a worm operator still commutes with
the interaction part $V$ of the Hamiltonian.  
By giving some worm components a bigger weight, the corresponding
components $G(r)$ will be updated more often, leading to a higher
accuracy and mimicking flat histograms \cite{Wang}.
The condensed
fraction $\rho_c$ can be determined via
\begin{equation}
\rho_c = \frac{1}{N N_s} \sum_{i,j}^{N_s} \langle b^{\dagger}_i b_j  \rangle.
\label{eq:cf}
\end{equation} 
The 
superfluid fraction can be determined using the winding number \cite{Ceperley87},
\begin{equation}
\rho_s = \frac{\langle W^2  \rangle N_s^2}{2tN\beta}, 
\label{eq:sf}
\end{equation}
where $\langle W^2  \rangle$ is the mean square of the winding number operator
in one dimension.
Figure \ref{fig:rho} shows the condensed and superfluid fraction for a
uniform one-dimensional system of 128 sites at a density of exactly one
particle per site.
Calculations were performed at an inverse temperature $\beta=128 t^{-1}$, using the
algorithm based on solution A.
We have used the algorithm to study the quantum critical behavior of the
one-dimensional Bose-Hubbard model with constant filling, using Renormalization
Group flow equations.
Studying the BKT transition is notoriously difficult
because of the logarithmic finite-size corrections.
The present algorithm has the big advantage of keeping the
density constant, in contrast to the grand-canonical approaches.  
The results of this study will be presented elsewhere \cite{Lodethesis}.

\begin{figure}[h]
\begin{center}
\includegraphics[angle=270, width=8.5cm]{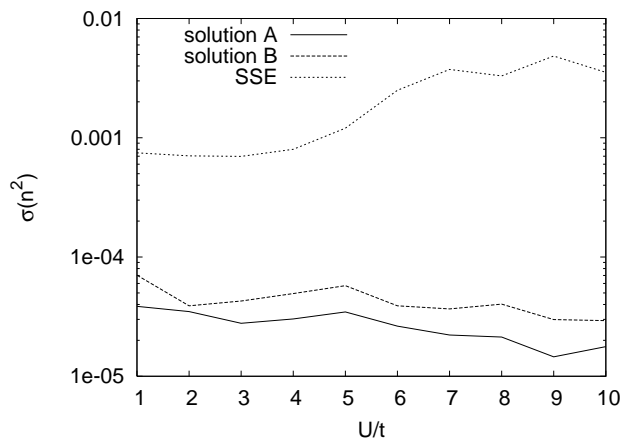}
\caption{A comparison between the 
 standard deviations on the squared density (see Eq. (\ref{eq:sqdens})) resulting
  from the directed loop  SSE algorithm and the
  algorithms based on solutions A and B. 
  The homogeneous Bose-Hubbard model is simulated for 
a lattice  with $32$ sites and $32$
  bosons at an inverse temperature $\beta=32 t^{-1}$. The deviations resulted
  after a QMC calculation with $40$ independent Markov chains that each ran $600$ seconds on a
  Pentium III processor.
   \label{fig:errornn2}}
\end{center}
\end{figure}

\begin{figure}[h]
\begin{center}
\includegraphics[angle=270, width=8.5cm]{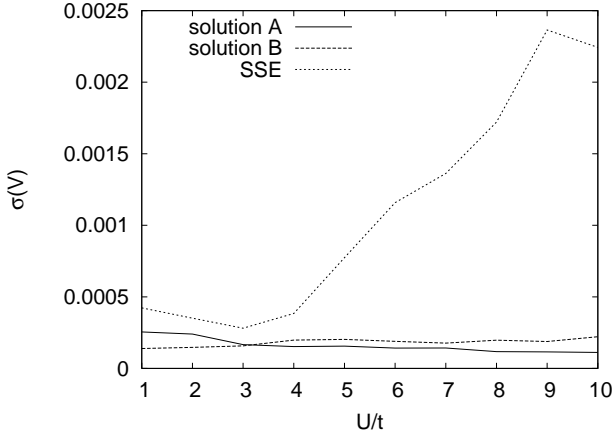}
\caption{The standard deviation on the mean value of $V$ (see
  Eq. \ref{eq:bhham}) for the homogeneous
  Bose-Hubbard model as a function of $U/t$. Here solution A is the most
  efficient one in the Mott phase. In the superfluid phase, solution B becomes
  more efficient.
   \label{fig:errorsen2}}
\end{center}
\end{figure}

\begin{figure}[h]
\begin{center}
\includegraphics[angle=270, width=8.5cm]{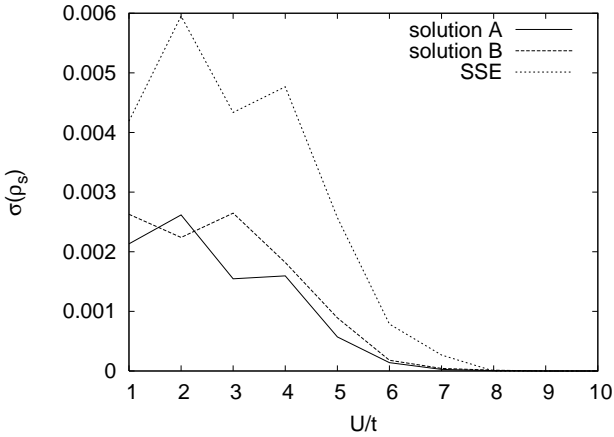}
\caption{The standard deviation on the mean value of the superfluid fraction
  $\rho_s$. The deviations result from solutions A and B, 
and the directed loop SSE method. Each simulation
  consisted of $40$ independent Markov chains that each ran $600$ seconds. 
   \label{fig:errorsf}}
\end{center}
\end{figure}

We compare the efficiencies of solutions A and B. To this purpose,
we simulate a one-dimensional lattice with $32$ sites at an inverse
temperature $\beta=32 t^{-1}$ and a filling factor of one boson per site. The simulations consisted of $40$ Markov chains that each ran
$600$ seconds after thermalization on a Pentium III processor. 
The same code was used, 
with only minor changes to go from algorithm A to B.
We discuss the efficiency by looking
at the standard deviations on the expectation value of $V$ of Eq. (\ref{eq:bhham})
and on the average squared density. We calculated the squared
density $n^2$ by averaging over all sites, 
\begin{equation}
\langle n^2 \rangle = \frac{1}{N_s} \sum_{i}^{N_s} \langle n_i^2  \rangle.
\label{eq:sqdens}
\end{equation}
The expectation value of the interaction term $V$ was not calculated via the
correlation function
$G(1)$, but
by counting the number of interaction vertices in the configuration whenever the
worm operator was diagonal \cite{Sandvik97}. When looking at the standard deviation on the
squared density (Figure \ref{fig:errornn2}), one can conclude solution A is the
most efficient one. We found a similar picture when looking at the standard
deviations on the expectation value of $H_0$ of Eq. (\ref{eq:bhham}).
The errors on the standard deviations lie within ten percent.
From the standard deviation on the expectation value of the interaction term $V$ (Figure
\ref{fig:errorsen2}), it follows that solution B is better in the
superfluid phase. The same conclusion follows from the total energy. 
For the condensed fraction the deviations are smallest for solution A for all values of $U/t$.
Those on the superfluid
fraction  lie very close for solution A and B (see Figure \ref{fig:errorsf}). 
We found that varying the parameter $\phi$ of Tables \ref{table:solA} and
\ref{table:solB} does not change the efficiency in a
significant way, as long as $\phi$ is not too small.
For further simulations we will always choose the parameter $\phi$ as big as possible, under the constraint $\phi
\leq \mathcal{N}_{DD'}$. 
Figures \ref{fig:errornn2}, \ref{fig:errorsen2} and
\ref{fig:errorsf} show 
also standard deviations resulting from  the
directed loop SSE algorithm \cite{Syl02, Syl03, Pollet04}.
One has to be very careful when comparing efficiencies of 
different algorithms. First, the SSE code works in the
grand-canonical ensemble. 
In the SSE simulations, the chemical potential was changed in such a way that
$N$ remained constant.
Second, the efficiency does not only depend on the
algorithm, but also on the used data structures.
In a SSE approach, the decomposition of the evolution operator corresponds to
a perturbation expansion in all terms of the Hamiltonian, while the
decomposition Eq. (\ref{eq:decompos}) perturbs only in the off-diagonal terms
$V$.
For the Bose-Hubbard model, where the contribution of the diagonal terms is
large, the last approach is preferable.
 For all calculated observables the standard deviations resulting from
the SSE code were the largest. Figures \ref{fig:errornn2} and
\ref{fig:errorsen2} show the SSE deviations increase rather rapidly with
increasing $U/t$, whereas the deviations resulting from our method 
remain of the same order. 
We also calculated autocorrelation times for different observables.
Here each bin ended after a constant number of measurements. 
We noticed that for solution B the autocorrelation times became
very big for high values of $U/t$.  For small $U/t$, the autocorrelation time
for solution A is of the order of the number of Markov steps needed for
$10$  diagonal updates, and increases only slowly with
increasing $U/t$. Of course it should be noted the algorithm based on solution
A had to run much longer in order to get the same number of diagonal measurements.
For all measured observables we found similar autocorrelation times.  

We conclude that solution A, derived in the previous section, is more
efficient than solution B, except in the superfluid phase when looking at the
interaction energy $V$.  This can be understood by remarking that the time
shifts of the worm operator are much larger for solution A. In the algorithm
based on solution B, the time shifts are of the order of the mean imaginary
time between two interaction vertices in the configurations. This explains 
why the standard deviations on the interaction energy are
smallest for this solution in the superfluid phase.
We also conclude that our algorithm is more efficient than the directed loop SSE
algorithm when simulating the one-dimensional Bose-Hubbard model. 
In the next section, we will apply the algorithm to a pairing model. In what
follows, all calculations are performed using the algorithm based on solution B.

\section{Application to a pairing model } 

In the nuclear shell
model, quantum Monte Carlo methods are valuable  
because they offer the possibility of  
doing calculations in much larger model spaces than conventional diagonalization
techniques. 
Finite temperature shell model studies have been done with the aid of auxiliary-field QMC
methods \cite{Koonin97},  and ground-state properties of light nuclei have been calculated
using variational and diffusion QMC techniques \cite{Pudliner95}.
Furthermore, being able to calculate thermal properties of nuclei makes it in
principle possible to calculate nuclear level densities
\cite{Nakada97}.
These densities are extremely important for making good theoretical estimates of
nuclear reaction rates. 

The basic assumption in the shell model is the presence of a mean field in
which the nucleons move. To improve on this, a residual interaction between
the nucleons is introduced. Pairing between nucleons is the main
short-range correlation induced by the residual interaction.
Adding a simple pairing Hamiltonian to the mean-field Hamiltonian
\begin{equation}
H_{mf} + H_P = \sum_{t=p,n}\sum_{k} e_{kt} n_{kt} - \sum_{t=p,n} \frac{G_t}{4} \sum_{k, k'} a^{\dagger}_{k't}
a^{\dagger}_{\overline{k'}t} a_{\overline{k}t} a_{kt}, 
\label{eq:pairingham}
\end{equation}
 can account for this \cite{Ring80, Dean03}. The operators $a^{\dag}_{kt}$ create a particle in the
 single-particle eigenstate $k$ of the mean-field Hamiltonian in the valence
 shell. 
The index $t$ indicates proton or neutron states and $\overline{k}$ denotes the
 time-reversed state of $k$. 
The operator $n_{kt}$ is the corresponding number-operator and
 $e_{kt}=e_{\overline{k}t}$ the single particle energy-eigenvalue. 
$G_t$ is the pairing strength for protons or neutrons.
Proton-neutron pairing is
 not included, but this coupling contributes only in an important way for $N=Z$ nuclei \cite{Langanke97}. 
As a consequence, the problem decouples for protons and
 neutrons. In the sequel only the neutron part of the model is considered, and 
 the isospin index $t$ is dropped to ease the notations.

Based on algebraic techniques developed by Richardson \cite{Richardson}, the
pairing model can be solved exactly for an arbitrary set of single particle levels at zero
temperature \cite{Rombouts04}.
In practice it remains difficult to use these exact results to study the
thermodynamics of the model, because the number of states needed in the
ensemble increases very rapidly with increasing temperature.    
Thermodynamical properties have been studied using 
auxiliary-field QMC, which is free of sign problems for the pairing Hamiltonian
of Eq. (\ref{eq:pairingham}) when dealing with an even number of particles \cite{Rombouts98}. 
However, the present algorithm can
consider nuclei with even and odd nucleon numbers. Note the auxiliary-field
method scales as O$(N_s^3)$ with $N_s$ the number considered single particle
states, while a world-line algorithm scales linear with $N_s$.

When a nucleon occupies a single particle state $k$ and its time-reversed state
$\overline{k}$ is unoccupied, the nucleon is said to be 'unaccompanied'.
These states do not
participate in the pair scattering by $H_P$. The mean-field plus pairing
Hamiltonian can be rewritten as Eq. (\ref{eq:hamiltonian}), 
 \begin{eqnarray}
  H & = & H_0 - V, \\
  H_0 & = & \sum_{k} e_{k} n_k - \frac{G}{2} \sum_{k}
  b^{\dagger}_{k} b_{k}, \\
  V & = & \frac{G}{4} \sum_{k\neq k'} b^{\dagger}_{k} b_{k'}. \label{eq:Vpairing}  
\end{eqnarray}
 The operators
 $b^{\dagger}_{k} = a^{\dagger}_{k}
a^{\dagger}_{\overline{k}}$ create a pair of nucleons in two
time-reversed states   
and satisfy hard-core boson commutation relations.
In order to get the correct finite temperature properties, 
the possibility of changing the number of 
unaccompanied nucleons during the simulation should be incorporated.
A path integral Monte Carlo method for the pairing Hamiltonian has been developed
by Cerf and Martin \cite{Cerf93, Cerf96}, but there the number  of pairs remained
fixed \cite{Romboutscom, Rombouts98}.
This problem can now be overcome elegantly by adding an extra pair breaking term
\begin{equation}
V_{\rm{pert}} = \frac{G g}{2}
\sum_{k} \sum_{k'\neq k''} \big( b^{\dagger}_k
a_{k'} a_{k''} + H.c.
\big),
\label{eq:Vpert}
\end{equation}
to the interaction part $V$ of Eq. (\ref{eq:Vpairing}).  
We define the worm operator as
\begin{equation}
A = \frac{1}{\bar{N}} \sum_{k} n_{k}
+ \frac{1}{4} \sum_{k\neq k'} b^{\dagger}_k b_{k'} + \frac{g}{2}
\sum_{k} \sum_{k'\neq k''} \big( b^{\dagger}_k
a_{k'} a_{k''} + H.c.
\big),
\label{eq:wormBCS}
\end{equation}
with two extra parameters $\bar{N}$ and $g$ to be optimized.
A term proportional to $V_{\rm{pert}}$ is included in the worm operator, in order
to satisfy condition Eq. (\ref{eq:commutator}).
This term will generate configurations with pair breaking interactions.
However, it can occur that too many of these
interactions are generated, though we are only
interested in generating configurations with a different number of unaccompanied
particles, but without
interactions of the type Eq. (\ref{eq:Vpert}). This can be prevented by imposing
the constraint that  a configuration can contain at most two pair breaking
interactions of this type. 
Observables are only updated if there are no pair-breaking interactions in the
configuration. 
This means that a number of Markov steps are needed in order to
reach a new allowed configuration with a different number of unaccompanied particles.
When $g$ of Eq. (\ref{eq:wormBCS}) is put equal to one, the percentage of diagonal
configurations which contain no $V_{\rm{pert}}$ interactions  
(see Eq. (\ref{eq:Vpert})) is about $15 \% $.  This is still efficient enough to sample
the pairing Hamiltonian. There are a number of ways to increase the efficiency.  First of all one can change the parameter $g$, hereby
influencing the appearance of pair breaking interactions. 
One can also restrict the number of times the
worm tries to insert a $V_{\rm{pert}}$ interaction by
allowing this only after a certain
Markov time in which 'good'
(i.e. without pair breaking interactions) configurations are sampled. One should also keep in mind that while a configuration contains
pair breaking interactions, the worm itself is not necessarily of the pair breaking
type.   So a lot of Markov time is spend to change the configuration in a
global way
without removing the pair breaking interactions, leading to strong decorrelation.

\begin{figure}[h]
\begin{center}
\includegraphics[angle=270, width=8.5cm]{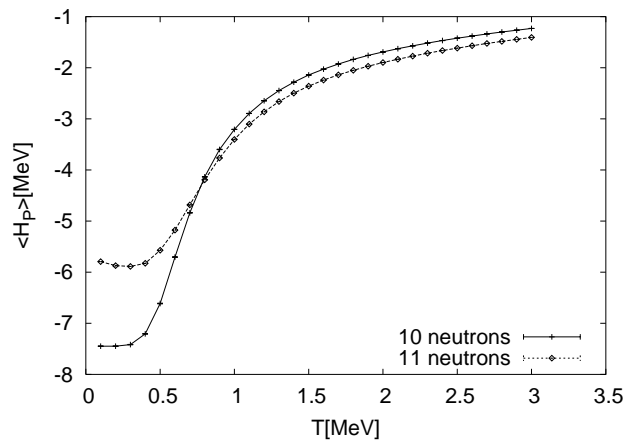}
\caption{Expectation value of the neutron part of the pairing-interaction
  operator as a function of temperature. The pairing strength $G_n$ is equal to
  $16/56$ MeV. We consider $10$ and $11$ neutrons in the $fp+sdg$ valence
  space (see Tabel \ref{table:spe}). At temperatures below $0.5$ MeV the
  pairing energy is much lower for the even neutron number.
   \label{fig:pairingn}}
\end{center}
\end{figure}

\begin{figure}[h]
\begin{center}
\includegraphics[angle=270, width=8.5cm]{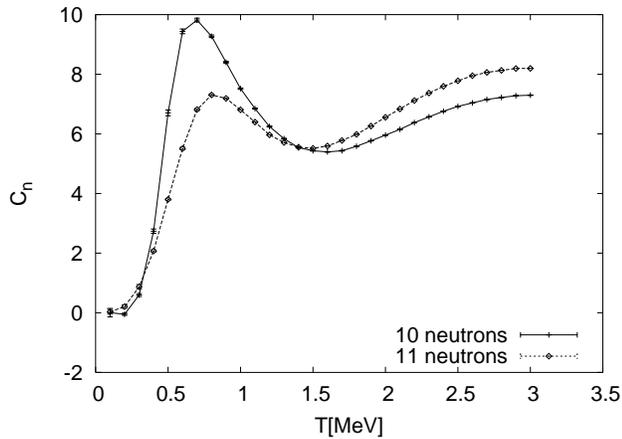}
\caption{ The neutron specific heat $C_n$ as a function of temperature for
  $10$ and $11$ neutrons in the full $fp + sdg$ valence space (see Table \ref{table:spe}).
Calculations were performed at a constant neutron pairing strength $G_n = 16/56$ MeV.
   \label{fig:shn}}
\end{center}
\end{figure}

The main physical properties of nuclei in the Iron region are modeled by a
schematic mean-field plus pairing Hamiltonian.
For the mean-field potential, we use a Woods-Saxon potential. Single particle
energies are taken from Ref. \cite{Rombouts98}. A full $fp+sdg$ valence space
is chosen. These single-particle states and energies are shown in Table \ref{table:spe}. 
A pairing strength $G = 16/56$ MeV is used. Due to the size of the model
space a strength smaller than the suggested value of $20$ MeV per nucleon is used \cite{Rombouts98}.
We have tested our code by comparing finite temperature results in a
$fp$
valence space with the ones obtained via an exact diagonalization technique \cite{Rossig96}.
We show results of calculations with the valence shell given in Table \ref{table:spe} occupied
by $10$ and $11$ valence neutrons.   
Figure \ref{fig:pairingn} shows the expectation value of the neutron pairing-interaction operator
$\langle H_P  \rangle$ as a function of temperature. 
At low temperature, the pairing energies are much lower for the even number of
neutrons. This can be understood by remarking that for an odd
number of neutrons there is always at least one unpaired
nucleon. At temperatures higher than $1$ MeV, the pairing energies differ only slightly,
because there is an increasing number of unpaired nucleons due to thermal excitation.
This is also reflected in the specific heat (see Figure \ref{fig:shn}).
A peak appears around $0.8$ MeV due to the development of pair correlations.

\begin{table}
\begin{tabular*}{0.4\textwidth}{@{\extracolsep{\fill}} |r|r|r|}
\hline
\hline
\multicolumn{3}{|c|}{Single-particle energies (MeV)} \\
\hline
Orbital &  Protons & Neutrons \\
\hline
$1f_{7/2}$ &  -4.1205 & -10.4576 \\
$2p_{3/2}$ &  -2.0360 &  -8.4804 \\
$2p_{1/2}$ &  -1.2334 &  -7.6512 \\
$1f_{5/2}$ &  -1.2159 &  -7.7025 \\
$3s_{1/2}$ &   4.7316 &  -0.3861 \\
$2d_{5/2}$ &   5.6562 &   0.2225 \\
$2d_{3/2}$ &   6.1324 &   0.9907 \\
$1g_{9/2}$ &   6.6572 &   0.5631 \\
\hline
\hline
\end{tabular*}
\caption{\label{table:spe} Single particle eigenstates of a Woods-Saxon
  potential, taken from Ref. \cite{Rombouts98}. The chosen valence space
  contains $42$ states. The proton and neutron single particle energies (in
  MeV) are shown on the right.}
\end{table}

Because the worm operator conserves angular momentum, one can restrict the
intermediate states to a specific value of the quantum numbers $J$ and $J_z$.
This is not possible with the auxiliary-field QMC method.
In our current implementation of the algorithm however, 
the occupation of each couple of time-reversed single-particle states $(k,
\overline{k})$ is exactly known at all times.
Because the unaccompanied particle number
operator
\begin{equation}
N^u = \sum_{k} \big( n_k -
  b^{\dagger}_{k} b_{k} \big), \\
\label{eq':unaccnumber}
\end{equation}
commutes with the angular momentum projection operator $J_z$ (but not with
$J^2$), our current code allows 
restricting the simulation to configurations with a fixed $J_z$.
Work on extending this technique to full J-projection is in progress.

\begin{figure}[h]
\begin{center}
\includegraphics[angle=270, width=8.5cm]{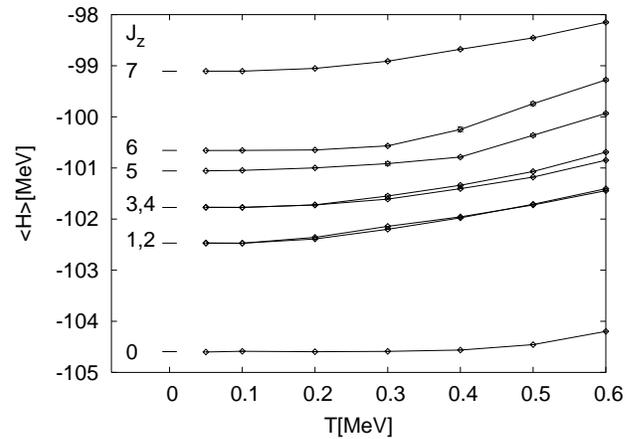}
\caption{ $J_z$ projected internal energies as a function of
  temperature. The values of $J_z$ from $0$ to $7$ are indicated on the left.
A clear convergence to 
exact zero temperature results calculated via techniques explained
  in Ref. \cite{Rombouts04} can be seen.
   \label{fig:jzproj}}
\end{center}
\end{figure}

When the projection on $J_z$ was turned on, we included an extra global
step in order to get a good convergence at the lowest temperatures. 
This extra global change 
allows for one or two
unaccompanied nucleons (which block the state they occupy) to move to
other states, and 
can occur whenever the worm is diagonal. 
First an
unaccompanied nucleon at a blocked state $l$ is chosen at random.
A 'non-blocked' pair of 
states $(k, \overline{k})$ is then chosen with probability
\begin{equation}
  P(k) =  e^{\int_0^{\beta}(n_k(t)+n_{\overline{k}}(t)-1)e_k dt} /N_l, 
\label{eq:probunac}
\end{equation}
with $n_k(t)$ the occupation number of state $k$ at imaginary time $t$, and
$N_l$ a normalization factor. The subscript $l$ indicates that the norm is
determined for a configuration containing a blocked state $l$.
 The idea behind Eq. (\ref{eq:probunac}) is to get a probability distribution $P(k)$
 which is peaked around the Fermi level, but other distributions can be chosen
 as well.
The interchange of the occupations of the blocked pair of states $(l,
\overline{l})$ and the non-blocked pair $(k, \overline{k})$
over the whole imaginary time interval $\beta$, 
is accepted with probability 
\begin{equation}
p = {\rm min}(1, \frac{N_l}{N_k}).
\end{equation}
The acceptance factor for the case when 
the occupations of two pairs of non-blocked and blocked states 
are interchanged,
can be
constructed in a similar way.  
The extra step has a high acceptance rate, but is only necessary to enhance decorrelation
at very low temperature when a $J_z$-projection is included. At higher
temperatures the unaccompanied nucleons move efficiently from state to state
via the last worm piece in Eq. (\ref{eq:wormBCS}). 
Figure \ref{fig:jzproj} shows total energies after $J_z$-projection at low
 temperature. Calculations were performed for ten neutrons moving
in the model space listed in Table \ref{table:spe}. The neutron pairing
strength is again $G_n = 16/56$ MeV.   
The figure also shows exact energy eigenvalues for  $J_z=0$ to $J_z=7$. 
These were calculated via a technique explained in Ref. \cite{Rombouts04}. 
The lowest $J_z=1,2$ and the lowest $J_z=3,4$ states are degenerate.
For low enough values of $T$
the finite temperature results clearly  converge to the ground states within the considered
ensembles. 

Note that we can compare with exact solutions because the pairing
strength was taken constant for all levels. Our QMC method allows to 
solve pairing models with a 
single particle state dependent pairing strength $G_{kk'}$, for which no algebraic solutions are
available. Taking in mind the method is applicable for even and odd nucleon
systems and allows angular momentum symmetry projections, this could greatly
extend the applicability of the pairing model.

\section{Conclusions and outlook } 

We have set up a  quantum Monte Carlo method with a non-local loop
updating scheme starting from a local worm operator in the path integral approach. This method  allows to sample
configurations with specific symmetries and, in particular, to sample the
canonical ensemble. It leads to a very efficient sampling scheme with all
moves accepted and without 'bounces' or critical slowing down near second
order phase transitions. We have proven
detailed balance and tested ergodicity. 
Our method opens new perspectives for the study of quantum many-body
systems where particle number and other symmetries play an important role.
It can be
applied to bosons, to fermions in absence of a sign problem and to
non-frustrated spin
systems at fixed magnetization. 
We have demonstrated this by simulating the Bose-Hubbard model and a nuclear
pairing model.
The equal-time one-body Green's function can be evaluated with high
efficiency. 
When non-equal time observables are required, the current method can in
principle still be combined with conventional non-local worm steps. 
There is still a lot of freedom in choosing the algorithm parameters, which can
be used to optimize the algorithm.
For the Bose-Hubbard model we compared the efficiency of our algorithm (with different parameter sets)
with a directed loop SSE code. Though one should always be careful when
comparing different algorithms, we have strong indications that our method is very
efficient.
We have simulated a pairing model for even and odd particle numbers. 
Our finite temperature results clearly 
supplement algebraic methods and other QMC methods. 
Furthermore, a projection on angular momentum symmetries can be included. We have
demonstrated this by showing $J_z$-projected results. A work on full
$J$-projection is in progress.

The authors wish to thank K.Heyde, J. Dukelsky, S. Wessel, M. Troyer and
S. Trebst for
interesting discussions and the Fund for Scientific Research - Flanders
(Belgium), the Research Board of the University of Ghent and N.A.T.O. for
financial support.

\end{document}